\documentclass[aps,nofootinbib,showpacs,preprintnumbers,amsmath,amssymb]{revtex4}
\def\be{\begin{equation}}
\def\ee{\end{equation}}
\def\bea{\begin{eqnarray}}
\def\eea{\end{eqnarray}}
\newcommand{\A}{{\mathcal{A}}}
\newcommand{\tA}{{\widetilde {\mathcal{A}}}}
\newcommand{\ta}{{\widetilde a}}
\newcommand{\tal}{{\widetilde \alpha}}
\newcommand{\tu}{{\widetilde u}}

\newcommand{\td}{{\widetilde d}}
\newcommand{\tQ}{{\widetilde Q}}

\usepackage{graphics}
\usepackage{graphicx}
\usepackage{dcolumn}
\usepackage{bm}
\usepackage{epsfig}
\usepackage{graphicx}
\usepackage{multirow}

\begin{document}

\vspace{1cm}

\preprint{USM-TH-291; arXiv:1107.2902v2}

\begin{flushright} {\bf} \end{flushright}
 
\title{Applying generalized Pad\'e approximants in analytic QCD models}

\author{Gorazd Cveti\v{c}}
 \email{gorazd.cvetic@usm.cl}
\affiliation{Department of Physics and Centro Cient\'{\i}fico-Tecnol\'ogico de Valpara\'{\i}so,
Universidad T\'ecnica Federico Santa Mar\'{\i}a, Casilla 110-V, 
Valpara\'{\i}so, Chile}

\author{Reinhart K\"ogerler} 
 \email{koeg@physik.uni-bielefeld.de}
\affiliation{Department of Physics, Universit\"at Bielefeld, 33501 Bielefeld, Germany}
 
\date{\today}

\begin{abstract}
A method of resummation of truncated perturbation series,
related to diagonal Pad\'e approximants but giving results 
independent of the renormalization scale, 
was developed more than ten years ago by us with a view of applying
it in perturbative QCD. We now apply this method in analytic QCD models,
i.e., models where the running coupling has no unphysical singularities, and 
we show that the method has attractive features such as a rapid convergence.
The method can be regarded as a generalization of the scale-setting methods of 
Stevenson, Grunberg, and Brodsky-Lepage-Mackenzie.
The method involves the fixing of various scales and weight coefficients via
an auxiliary construction of diagonal Pad\'e approximant. In low-energy
QCD observables, some of these scales become sometimes low at high order, 
which prevents the method from being effective in perturbative QCD 
where the coupling has unphysical singularities at low spacelike momenta. 
There are no such problems in analytic QCD.
\end{abstract}

\maketitle

\section{Introduction}
\label{sec:intr}  
Extending the applicability of QCD from high energies, where it can be
consistently  treated by perturbation methods, down to
the low-energy regime is one of the main tasks of theoretical hadronic
physics. A simple-minded utilization of perturbation series is clearly
forbidden, not just by the sheer size of the expansion parameter (the
running coupling parameter $a(Q^2) \equiv \alpha_s (Q^2)/\pi$ 
at low momentum transfer $Q^2 \equiv -q^2$), 
but even more so by the existence of unphysical (Landau) 
singularities of the coupling parameter in the complex $Q^2$ plane,
the singularities which are inferred from the renormalization group
equation when the corresponding beta function is expressed in terms 
of a truncated perturbation series. These singularities are
unphysical because they do not reflect correctly the analytic
properties of spacelike observables ${\cal D}(Q^2)$, properties
based on the general principles of local quantum field theories 
\cite{BS,Oehme}.
Consequently, the most straightforward procedure for applying QCD to 
low-energy quantities consists in removing this unwanted 
nonanalyticity by some kind of analytization of the coupling parameter
$a(Q^2) \mapsto \A_1(Q^2)$. The analytic coupling parameter
$\A_1(Q^2)$ can differ significantly from the perturbative one
$a(Q^2)$ only at low momenta $|Q^2| \alt 1 \ {\rm GeV}^2$.
Several constructions of such analytic QCD models, i.e., of $\A_1(Q^2)$,
have been made during the last fifteen years -- 
starting from the seminal papers of Shirkov et al.~\cite{ShS,MSS,Sh}. 
For reviews of various types of analytic QCD models see 
Refs.~\cite{Prosperi,Shirkov,Cvetic,Bakulev}.
On the other hand, handling the physics of hadrons at low energies by
simply utilizing an appropriately modified, ``analytized,'' coupling
parameter (together with its higher order analogs) within 
perturbative approaches is a very ambitious task, since it implicitly rests
on the assumption that the low-$Q^2$ behavior of $\A_1(Q^2)$ can be
defined in a way that all nonperturbative effects are effectively
included -- at least for inclusive quantities. Of particular interest here
is the behavior of $\A_1(Q^2)$ for $Q^2 \to 0$, and this question
was the subject of intensive studies during last years, 
based either on analytic methods
(Schwinger-Dyson equations \cite{SDEs}, Banks-Zaks expansion \cite{BZ,Gardi:1998qr})
or on numerical lattice approaches \cite{latt}.
They have finally led to the strong
suspicion of ``freezing'' of the coupling parameter near $Q^2 = 0$. 
If one wants to go a step further, however, and specify 
$\A_1(Q^2)$ for the whole range $|Q^2| \alt Q^2_{as}$ 
($Q^2_{as}$ denotes the momentum transfer where
asymptotic freedom should start to dominate) such that all non-perturbative 
effects get included, one clearly has to utilize as much as possible 
external information, both on the side of
empirical constraints and on the side of general physical principles such
as causality, unitarity, analyticity, asymptotic freedom, operator product
expansion, renormalization scale and scheme independence, etc.

Within the present paper we focus mainly on the analytical structure and 
on  the renormalization scale (RScl) independence of the resulting physical 
quantities. We apply, in various analytic QCD models, a global 
(i.e., nonpolynomial in the coupling) RScl invariant resummation/evaluation
method which we developed in the context of perturbative QCD
more than ten years ago \cite{GC,GCRK}, and we compare this
evaluation method with other methods.
In Sec.~\ref{sec:recap} we recapitulate the aforementioned RScl invariant
resummation method for spacelike observables (in perturbative QCD).
The presentation this time is
somewhat less formal and, perhaps, more intuitive.
In Sec.~\ref{sec:appanQCD} we describe the minimal adjustments needed
for the method to be used in analytic QCD models. In that Section
we also argue why we should expect our resummation method to
work significantly better in analytic QCD than in perturbative QCD.
In Sec.~\ref{sec:num} we apply the method to the evaluation of the
derivative of the massless (vector) current-current correlation function,
i.e., the Adler function, both in perturbative QCD and in
various motivated analytic QCD models. First, the evaluations
are made for the leading-$\beta_0$ part of the Adler function, where
we know the exact result within each analytic QCD model, so this case
is used as a test case for our resummation method to rather high
values of the order index $M$. Subsequently, we apply our method
to the truncated series of the Adler function, where only the
first three full coefficients (beyond the leading term) are known.
In Sec.~\ref{sec:concl} we summarize the results and present conclusions.

\section{Recapitulation of the method}
\label{sec:recap}

In this Section we present the resummation method developed
in Refs.~\cite{GC,GCRK} in a somewhat simpler and, perhaps, 
more intuitive way. We consider a massless
spacelike physical observable ${\cal D}(Q^2)$ whose perturbation
series in powers of the perturbative QCD (pQCD) coupling 
$a(Q^2) \equiv \alpha_s(Q^2)/\pi$  
\be
{\cal D}(Q^2)_{\rm pt} = a(Q^2) + 
\sum_{j=1}^{\infty} d_j \; a(Q^2)^{j+1} 
\label{Dptmu2Q2}
\ee
is known up to $\sim a^{2 M}$, such that we are faced with the truncated 
perturbation series ${\cal D}(Q^2)^{[2 M]}_{\rm pt}$ 
\be
{\cal D}(Q^2)_{\rm pt}^{[2 M]} = a(Q^2) + 
\sum_{j=1}^{2 M - 1} d_j \; a(Q^2)^{j+1} \ .
\label{Dtptmu2Q2}
\ee
Here we have chosen the renormalization scale (RScl) $\mu^2$ to be
equal to the physical scale $Q^2$ of the process ($\mu^2=Q^2$). For
a general RScl $\mu^2$, the full and the truncated perturbation series read
\bea 
{\cal D}(Q^2)_{\rm pt} &=& a(\mu^2) + 
\sum_{j=1}^{\infty} d_j(\mu^2/Q^2) \; a(\mu^2)^{j+1} 
\label{Dpt}
\\
{\cal D}(Q^2;\mu^2)_{\rm pt}^{[2 M]} &=& a(\mu^2) + 
\sum_{j=1}^{2 M - 1} d_j(\mu^2/Q^2) \; a(\mu^2)^{j+1} \ .
\label{Dtpt}
\eea 
This truncated series has a residual RScl dependence due to truncation.
The $\mu^2$-dependence of $d_j(\mu^2/Q^2)$ is dictated by the
$\mu^2$-independence of the full series ${\cal D}(Q^2)_{\rm pt}$ and the
$\mu^2$-dependence of $a(\mu^2)$ given by the well known
renormalization group equation
\be
\frac{d a(\mu^2)}{d \ln \mu^2} = 
- \sum_{j \geq 2} \beta_{j-2} \; a(\mu^2)^j = 
-\beta_0 a(\mu^2)^2 \, \left( 1+ c_1 a(\mu^2) +c_2 a(\mu^2)^2 + \dots \right)\
\ ,
\label{RGE}
\ee
where the right-hand side is the beta function $\beta(a)$,
and we denoted $c_j \equiv \beta_j/\beta_0$.
In particular, we obtain (we denote throughout: $d_j(1) \equiv d_j$ and
$d_0=d_0(\mu^2/Q^2) = 1$)
\bea
d_1(\mu^2/Q^2) &=& d_1 + \beta_0 \ln(\mu^2/Q^2) \ ,
\label{d1RScl}
\\
d_2(\mu^2/Q^2) & = & d_2 + \sum_{k=1}^2 \frac{2!}{k! (2-k)!} \; 
\beta_0^k \; \ln^k \left( \frac{\mu^2}{Q^2} \right) d_{2-k} + 
\beta_1 \ln \left( \frac{\mu^2}{Q^2} \right) \ ,
\label{d2RScl}
\\
d_3(\mu^2/Q^2) & = & d_3 + \sum_{k=1}^3 \frac{3!}{k! (3-k)!} \; 
\beta_0^k \; \ln^k \left( \frac{\mu^2}{Q^2} \right) d_{3-k} + 
\beta_1 \left[ 2 d_1 \ln \left( \frac{\mu^2}{Q^2} \right)
+ \frac{5}{2} \beta_0  \ln^2 \left( \frac{\mu^2}{Q^2} \right) \right] 
+ \beta_2 \ln \left( \frac{\mu^2}{Q^2} \right) \ ,
\label{d3RScl}
\eea
etc. Note that $a(\mu^2)$ and $d_j(\mu^2/Q^2)$ are not only 
renormalization scale (RScl) dependent, but also renormalization 
scheme (RSch) dependent (as are also $d_j \equiv d_j(1)$), i.e.,
they are functions of $\mu^2$, $c_2=\beta_2/\beta_0$, $c_3=\beta_3/\beta_0$, etc.
The RSch dependence of $d_j(\mu^2/Q^2)$ and $d_j$ involves $c_2,\ldots, c_j$
(when $j\geq 2$).
The first two coefficients $\beta_0$ and $\beta_1$ are universal in the mass independent schemes: $\beta_0=(11 - 2 n_f/3)/4$, $\beta_1=(102 - 38 n_f/3)/16$.
 
In the following we will mainly be interested in the RScl dependence 
of the different (perturbation) series. Therefore,
it will prove advantageous to use logarithmic derivatives of the
pQCD coupling $a$ instead of powers $a^n$. 
Specifically, we introduce\footnote{Note that the factor  
in front of the right-hand side is 
chosen such that ${\ta}_1 \equiv a$ and  
$\ta_{n+1} = a^{n+1} + {\cal O}(a^{n+2})$ for $n \geq 1$. 
Only at one-loop level
approximation we have $\ta_{n+1} = a^{n+1}$, 
but in general $\ta_{n+1} \not= a^{n+1}$.}
\be
{\ta}_{n+1}(Q^2)
\equiv \frac{(-1)^n}{\beta_0^n n!}
\frac{ d^n a(Q^2)}{d (\ln Q^2)^n} \ .
\label{tan}
\ee
and reorganize the (truncated) perturbation series (\ref{Dpt})-(\ref{Dtpt}) 
into the ``modified (truncated) perturbation series'' (mpt)
\bea
{\cal D}(Q^2)_{\rm mpt} &=& a(\mu^2) + 
\sum_{j=1}^{\infty} {\td}_j(\mu^2/Q^2) \; {\ta}_{j+1}(\mu^2) \ ,
\label{Dmpt}
\\
{\cal D}(Q^2; \mu^2)_{\rm mpt}^{[2 M]} &=& a(\mu^2) + 
\sum_{j=1}^{2 M - 1} {\td}_j(\mu^2/Q^2) \; {\ta}_{j+1}(\mu^2) \ .
\label{Dtmpt}
\eea
Here the coefficients ${\td}_j(\mu^2/Q^2)$ are chosen so that the expressions
(\ref{Dpt}) and (\ref{Dmpt}) are formally identical. The advantage of
using here the logarithmic derivatives (\ref{tan}) and the expansions
(\ref{Dmpt}) and (\ref{Dtmpt}),\footnote{
The logarithmic derivatives of the coupling and the expansions of the type
(\ref{Dmpt}) and (\ref{Dtmpt}) were used systematically in 
Refs.~\cite{GCCV1,GCCV2} (in the context of analytic QCD), and in
Ref.~\cite{CLMV} (in the context of pQCD).}
as opposed to the expansions (\ref{Dpt}) and (\ref{Dtpt}), 
lies principally in the simple recursion relations for $\ta_n$'s
\be
\frac{d}{d \ln \mu^2} \ta_n(\mu^2) = - \beta_0 n \ta_{n+1} \ ,
\label{rectan}
\ee
whereas for the powers $a^n$ the relation is more complicated
\be
\frac{d}{d \ln \mu^2} a(\mu^2)^n = - n \beta_0 a(\mu^2)^{n+1}
\left( 1 + c_1 a(\mu^2) + c_2 a(\mu^2)^2 + \cdots \right) \ ,
\label{recan}
\ee
the right-hand side here being the consequence of the RGE (\ref{RGE}).
When we use the fact that the full series ${\cal D}(Q^2)_{\rm mpt}$ in
Eq.~(\ref{Dmpt}) is RScl independent
\be
\frac{d}{d \ln \mu^2} {\cal D}(Q^2)_{\rm mpt} = 0 \ ,
\label{RSclindDmpt}
\ee
we obtain a set of differential equations 
\be
\frac{d}{d \ln \mu^2} \td_n(\mu^2/Q^2) = n \beta_0 \td_{n-1} (\mu^2/Q^2) 
\qquad (n=1,2,\ldots) \ ,
\label{tdndiff}
\ee
whose integration gives (we denote throughout $\td_j(1) \equiv \td_j$ and
$\td_0=1$)
\be
{\td}_n(\mu^2/Q^2) = {\td}_n + \sum_{k=1}^n \frac{n!}{k! (n-k)!} 
\ \beta_0^k \ \ln^k \left( \frac{\mu^2}{Q^2} \right) {\td}_{n-k} \ .
\label{tdnmu}
\ee
We note that the relations (\ref{tdnmu}) for ${\td}_n(\mu^2/Q^2)$, 
in contrast to those for $d_n(\mu^2/Q^2)$ in 
Eqs.~(\ref{d1RScl})-(\ref{d3RScl}), do not involve any higher-loop
beta coefficients $\beta_j$ ($j \geq 1$). 
Therefore, it is suggestive to compare the situation with the one-loop 
limit of QCD (where $\beta_1=\beta_2=\ldots = 0$). In that limit the 
perturbative coupling, now denoted as $a_{1 \ell}(\mu^2)$, has the one-loop 
RGE running from a given value $a(Q^2)$ at the scale $Q^2$ to the scale $\mu^2$
\be
a_{1\ell}(\mu^2) = 
\frac{a(Q^2)}{1 + \beta_0 \ln(\mu^2/Q^2) \; a(Q^2)} \ .
\label{a1l}
\ee
Furthermore, in this case we have $\ta_{n+1,1\ell}(\mu^2) =
a_{1\ell}(\mu^2)^{n+1}$, where $\ta_{n+1,1\ell}(\mu^2)$ are the logarithmic
derivatives of $a_{1\ell}(\mu^2)$ analogous to Eq.~(\ref{tan}).

Consequently, if we define the (auxiliary) quantity
${\widetilde {\cal D}}(Q^2)$ via the following power series:
\be
{\widetilde {\cal D}}(Q^2)_{\rm pt} = a_{1\ell}(\mu^2) + 
\sum_{j=1}^{\infty} {\td}_j(\mu^2/Q^2) \; a_{1\ell}(\mu^2)^{j+1} \ ,
\label{tDpt}
\ee 
then Eqs.~(\ref{tdnmu}) represent the correct $\mu^2$ dependence of the
coefficients so as to ensure $\mu^2$ independence of the auxiliary
quantity ${\widetilde {\cal D}}(Q^2)$. Phrased differently, the auxiliary
quantity (\ref{tDpt}) is exactly invariant under the combined
RScl transformations
\be
\td_j \to \td_j(\mu^2/Q^2) \quad {\rm via \ Eq.~(\ref{tdnmu})} \ ,
\quad
a(Q^2) \to  a_{1\ell}(\mu^2)  \quad {\rm via \ Eq.~(\ref{a1l})} \ .
\ee
Note that Eq.~(\ref{a1l}) has the form of a homographic transformation.
The latter observation leads to an appropriate way for treating 
truncated series, 
which are in general $\mu^2$ dependent due to truncation,
in particular ${\widetilde {\cal D}}(Q^2;\mu^2)_{\rm pt}^{[2 M]}$
(we consider truncated series with an even number of terms).
Namely, it is well known in mathematics that the diagonal
Pad\'e approximants (dPA's), being ratios of two polynomials ($P_M$, $R_M$),
both of order $M$
\be
[M/M](x) = P_M(x)/R_M(X)
\label{dPAMM}
\ee
remain dPA's under the homographic transformation
\be
x \mapsto {\overline x} = x/(1 + K x) \ ,
\label{homograph}
\ee
(where $K$ is an arbitrary constant). This means that
\be
[M/M]({\overline x}) = {\cal P}_M(x)/{\cal R}_M(x) \ ,
\label{dPAMMbarx}
\ee
where ${\cal P}_M(x)$ and ${\cal R}_M(x)$ are again two polynomials both
of order $M$.
More explicitly, if $[M/M]_{\bar f}(x)$ is the dPA of a function
${\bar f}(x)$ whose Taylor expansion around $x=0$ exists
(${\bar f}(x) - [M/M]_{\bar f}(x) \sim x^{2 M+1}$), then there exists a function
$F$ ($\not= {\bar f}$) such that $[M/M]_{\bar f}({\overline x}) = [M/M]_F(x)$. 
As a consequence,
it can be shown that for any function $f$ (with Taylor expansion around $x=0$)
the following identity holds:\footnote{
We have: $f(x) - [M/M]_f(x) \sim x^{2 M+1}$, and ${\bar f}({\overline x}) -
[M/M]_{\bar f}({\overline x}) \sim {\overline x}^{2 M+1} \sim x^{2 M+1}$. Therefore, since
$f(x) = {\bar f}({\overline x})$ and $[M/M]_{\bar f}({\overline x}) = [M/M]_F(x)$, we obtain:
$[M/M]_F(x) - [M/M]_f(x) \sim x^{2 M+1}$. This implies $[M/M]_f(x) = [M/M]_F(x)$
(i.e., $[M/M]_f(x) = [M/M]_{\bar f}({\overline x})$, Eq.~(\ref{dPAinv})),
because the $[M/M](x)$ Pad\'e's are uniquely determined by the coefficients
of their expansion in powers $x^n$ for $n \leq 2 M$.}
\be
[M/M]_f(x) = [M/M]_{\bar f}({\overline x}) \ ,
\label{dPAinv}
\ee
where ${\overline x} = x/(1 + K x)$ and ${\bar f}({\overline x}) = f(x)$. In our case of
${\widetilde {\cal D}}(Q^2)_{\rm pt}$ and its expansion (\ref{tDpt}), we identify:
$x = a(Q^2)$, ${\overline x} = a_{1\ell}(\mu^2) = x/(1 + K x)$ [$K= \beta_0 \ln(\mu^2/Q^2)$;
$\mu^2 = Q^2 \exp(K/\beta_0)$], and ${\widetilde {\cal D}}(Q^2)_{\rm pt} = f(x) = {\bar f}({\overline x})$.
The latter identification holds because ${\widetilde {\cal D}}(Q^2)_{\rm pt} =
x + \sum_{j=1}^{\infty} \td_j x^{j+1} = {\overline x} + \sum_{j=1}^{\infty} \td_j(\mu^2/Q^2) {\overline x}^{j+1}$.
The identity (\ref{dPAinv}) means that dPA's of ${\widetilde {\cal D}}(Q^2)_{\rm pt}$
have exact independence of the RScl $\mu^2$. Stated differently,
when constructing dPA of expansion (\ref{tDpt}), it does not matter which
value of the RScl $\mu^2$ we use in (\ref{tDpt}).

This fact was noticed by Gardi \cite{Gardi}, who, as a result,
argued that the truncated perturbation series of the form (\ref{Dpt})
for physical observables ${\cal D}(Q^2)$ can be well approximated by dPA's because
the result is approximately RScl independent (i.e., it is exactly
RScl-independent when the RGE-running is approximated to be one-loop). 
Here we see that these considerations are valid without approximation 
for the (RScl-independent) auxiliary quantity
${\widetilde {\cal D}}(Q^2)$ which is defined via the power series (\ref{tDpt}). 
This is related with the fact that the RScl dependence of the
coefficients $\td_j(\mu^2/Q^2)$ as given by Eq.~(\ref{tdnmu}), although
involving only $\beta_0$ and no higher $\beta_j$ coefficients, is exact.
On the other hand, the RScl dependence of the original
coefficients $d_j(\mu^2/Q^2)$ appearing in the power series (\ref{Dpt})
is more complicated and involves (for $j \geq 2$) higher-loop beta coefficients
$\beta_k$ ($k \leq j-1$), as seen in Eqs.~(\ref{d1RScl})-(\ref{d3RScl}).
 
The dPA $[M/M]$ of ${\widetilde {\cal D}}(Q^2)$ has the general form
\be
[M/M]_{\widetilde {\cal D}}(a_{1\ell}(\mu^2)) = 
x \frac{1 + A_1 x + \cdots A_{M-1} x^{M-1}}{1 + B_1 x + \cdots + B_M x^M}
{\bigg |}_{x=a_{1\ell}(\mu^2)} \ .
\label{MM}
\ee
We rewrite it by applying a partial fraction decomposition of the fraction
on the right-hand side.\footnote{
In Mathematica \cite{Math8}, the command ``Apart'' achieves this.}
If we denote the $M$ zeros of the denominator polynomial 
$(1 +  B_1 x + \cdots + B_M x^M)$ by $-1/\tu_j$ ($j=1,\ldots,M$), we obtain
\be
[M/M]_{\widetilde {\cal D}}(a_{1\ell}(\mu^2)) =  
\sum_{j=1}^M \tal_j \frac{x}{1 + \tu_j x}{\bigg |}_{x=a_{1\ell}(\mu^2)} 
\ , 
\label{MMdecom1}
\ee
with appropriate ``weights'' $\tal_j$  ($j=1,\ldots,M$). Using Eq.~(\ref{a1l})
gives us finally
\be
[M/M]_{\widetilde {\cal D}}(a_{1\ell}(\mu^2)) = 
  \sum_{j=1}^M \tal_j \; a_{1\ell}(\tQ_j^2) \ , \quad 
{\rm where \ } \tQ_j^2 = \mu^2 \exp(\tu_j/\beta_0) \ ,
\label{MMdecom2}
\ee
i.e., we expressed $[M/M]_{\widetilde {\cal D}}$ as a weighted average of
one-loop running couplings defined at specific reference momentum values
(gluon virtualities) $\tQ_j^2$  ($j=1,\ldots,M$).\footnote{
In principle, $-1/\tu_j$'s (and thus $\tQ_j^2$'s) and $\tal_j$'s 
can be sorted into complex conjugate pairs and into real values. 
In Sec.~\ref{sec:num} we apply this approach to the massless Adler function
for which it turns out that all $\tQ_j^2$ and $\tal_j$ are real.}
Since, as argued, the expressions (\ref{MM})-(\ref{MMdecom2})
are exactly independent of the RScl chosen in the original series
(\ref{tDpt}), both the weights $\tal_j$ and the scales $\tQ_j^2$
are exactly independent of this RScl. 

This observation helps us find an analogous approximant for the
true observable ${\cal D}$ (or its truncated version ${\cal D}^{[2 M]}$).
By comparing Eq.~(\ref{Dmpt}) with (\ref{tDpt}), we are motivated to
define the following approximant:
\be
{\cal G}^{[M/M]}_{{\cal D}}(Q^2) =  
\sum_{j=1}^M \tal_j \; a(\tQ_j^2) \ ,
\label{dBG}
 \ee
i.e., we simply replace in the expression (\ref{MMdecom2}) the
one-loop running coupling $ a_{1\ell}(\tQ_j^2)$ by the exact
($n$-loop running, $n$ arbitrary) coupling parameter $a(\tQ_j^2)$.

The resulting approximant has two important properties:
\begin{enumerate}
\item
\label{RSclinv}
It is, by sheer construction, exactly RScl invariant (since
$\tal_j$ and $\tQ_j^2$ are independent of $\mu^2$);
\item
\label{appr}
It fulfills the approximation requirement
\be
{\cal D}(Q^2) - {\cal G}^{[M/M]}_{{\cal D}}(Q^2) = {\cal O}(\ta_{2 M+1})
= {\cal O}(a^{2 M+1})
\ ,
\label{dBGappr}
\ee   
i.e., it reproduces the first $2 M$ terms of the series (\ref{Dmpt})
and of the series (\ref{Dpt}).
It is relatively straightforward to show the latter fact, by expanding the
expression (\ref{dBG}) in terms of logarithmic derivatives (see the Appendix).
\end{enumerate}

An approximant of the type (\ref{dBG}) was originally introduced in
Ref.~\cite{GC}, based on more mathematical considerations. It was called
``modified Baker-Gammel approximant'' and interpreted as a particularly
clever resummation procedure for the physical observable ${\cal D}(Q^2)$.
In Ref.~\cite{GC}, also a more formal proof of the properties \ref{RSclinv}
(RScl invariance) and \ref{appr} (approximation property) was given.
The proof rested on choosing the kernel of the Baker-Gammel approximant
to be  $k(z,\tu) = f(\tu)/z$ where 
$z= a(\mu^2)$, $\tu = \beta_0 \ln(\tQ^2/\mu^2)$ and $f(\tu) = a(\tQ^2)$.\footnote{
For the conventional Baker-Gammel approximants, see for example part II 
of Ref.~\cite{Baker}. Exact RScl invariance of such constructions in 
the special case of the aforementioned kernel was apparently first shown
in Ref.~\cite{GC}.} 
Within the present paper we constructed the same approximant
(\ref{dBG}) in a more heuristic and physically motivated manner.

In Ref.~\cite{GCRK} we extended the construction of this approximant 
so as to be applicable also to the case when an even number
of coefficients $d_j$ ($j=1, \ldots, 2 M$)
are known in the expansion (\ref{Dpt}), and in
Refs.~\cite{applGC} the method was applied in pQCD.

We can interpret the form (\ref{dBG}) as a kind of extension of the
previously known scale-setting techniques (principle of minimal
sensitivity \cite{PMS}, effective charge method \cite{ECH} and 
related approaches \cite{ECHext}, and the
scale-setting of Brodsky-Lepage-Mackenzie \cite{BLM} and
its extensions \cite{BLMext1,BLMext2,BLMext3}) to several scales.
However, in the presented case these scales are not fixed by a
specific motivated prescription of scale-setting, but are rather
based primarily on the successes of diagonal Pad\'e approximants in
physics and on the additional requirement of refining the
approximate (one-loop) RScl invariance of the approximant to the
exact RScl invariance. These approximants are global, i.e., they go beyond
the polynomial form in $a$, and this is one of the
reasons why we expect them to include nonperturbative effects.

Also interesting to note is the connection of our approximant
(\ref{dBG}) with Neubert's resummation method \cite{Neubert}
which is defined by integration over the momentum flow within
the running coupling parameter and the connected momentum
distribution function $w_{\cal D}$
\be
{\cal D}^{\rm (LB)}(Q^2)_{\rm pt} =    
\int_0^{\infty} dt \; w_{\cal D}(t) a(t Q^2 e^{{\overline {\cal C}}}) \ .
\label{DNeub}
\ee 
Here, ${\overline {\cal C}} = -5/3$ if the ``${\overline {\rm MS}}$'' 
convention for the scale $\Lambda_{\rm QCD}$ is used. When expanding the
parameter $a(t Q^2 e^{{\overline {\cal C}}})$ around $a(\mu^2)$, it turns out that
this expression represents exactly the leading-$\beta_0$ part (LB) of
the ``modified perturbation expansion'' (\ref{Dmpt})
(cf.~Ref.~\cite{GCCV2}, and Eq.~(\ref{DLBexp}) later in the present paper). 
We see that our approximant (\ref{dBG}) is
equivalent to an approximation of the distribution function $w_{\cal D}(t)$ in 
the integrand in (\ref{DNeub}) in terms of the
weighted sum of delta functions
\be
w_{\cal D}(t) \approx \sum_{j=1}^M \tal_j \delta(t - t_j) \ ,
\label{sumdelt}
\ee
where the delta peaks are located at $t_j$'s such that
$t_j Q^2 e^{{\overline {\cal C}}} = \tQ_j^2$ ($j=1,\ldots,M$).

\section{Application to analytic QCD models}
\label{sec:appanQCD}

In general, the perturbative QCD coupling $a(Q^2)$ has
a cut in the complex $Q^2$ plane along the negative semiaxis
up to the positive Landau branching point $\Lambda_{\rm L}^2$. 
On the other hand, by the general
principles of the local and causal quantum field theory  \cite{BS,Oehme},
the spacelike observables ${\cal D}(Q^2)$ (such as the Adler function,
sum rules, etc.) must be analytic functions in the $Q^2$ complex plane
with the exception of the cut on the negative semiaxis 
$Q^2 \in \mathbb{C} \backslash (-\infty, 0]$. This analyticity property, however,
is not reflected by the $a(Q^2)$ which has a cut on a part of
the positive axis $[0, \Lambda_{\rm L}^2]$. Therefore, various analytic QCD
models have been constructed where the nonanalytic $a(Q^2)$
is replaced by an analytic $\A_1(Q^2)$ which has no singularities
for $Q^2 \in \mathbb{C} \backslash (-\infty, 0]$ and at high $|Q^2| \gg \Lambda^2$
(approximately) agrees with  $a(Q^2)$. 
For details on some of such models we refer to
various references: minimal analytic (MA) model \cite{ShS,MSS,Sh,BMS};
modified minimal analytic model \cite{Nesterenko}; analytic perturbative
models \cite{GCRKCV}; a specific (``close to perturbative'') analytic
model \cite{CCEM}. Reviews of analytic QCD models are given in 
Refs.~\cite{Prosperi,Shirkov,Cvetic,Bakulev}. 
Calculational techniques applicable to any 
analytic QCD model (the latter being defined via a specification of $\A_1(Q^2)$
only) are described in Refs.~\cite{GCCV1,GCCV2,GCAK}.

It is natural to ask: how do our approximants ${\cal G}^{[M/M]}$ fare in
such analytic QCD models. As mentioned above, these approximants 
(\ref{dBG}) choose specific scales which, for low-energy observables, 
are often close to or
inside the (unphysical) Landau singularity regime of $a(Q^2)$.
Therefore, the hope is that our approximants fare much better or
even develop all their potential in analytic QCD models where they
look simply as
\be
{\cal G}^{[M/M]}_{{\cal D}}(Q^2;{\rm an.}) =
\sum_{j=1}^M \tal_j \; \A_1(\tQ_j^2) \ .
\label{dBGan}
 \ee
The other intriguing aspect is that, in any analytic QCD model\footnote{
We regard the specification of the coupling function $\A_1(Q^2)$ in the
complex $Q^2$ plane as the full specification of an analytic QCD model.}
the analytization of the higher powers $a^n$ goes in fact via
the analytization of the logarithmic derivatives (\ref{tan}), 
cf.~Refs.~\cite{GCCV1,GCCV2}
\be
{\ta}_{n+1} \mapsto \tA_{n+1}
\quad (n=0,1,2,\ldots) \ ,
\label{anrule1}
\ee
where ${\tA}_{n+1}$ are the logarithmic derivatives
of the analytic coupling $\A_1$
\be
\tA_{n+1}(Q^2) \equiv \frac{(-1)^n}{\beta_0^n n!}
\frac{ \partial^n \A_1(Q^2)}{\partial (\ln Q^2)^n} \ ,
\qquad (n=0,1,2,\ldots) \ ,
\label{tAn}
\ee
and not via the naive replacement $a^n \mapsto \A_1^n$.\footnote{
The analytic analogs $\A_n(Q^2)$ of powers $a(Q^2)^n$ are obtained from
the relations $\A_n = \tA_n + \sum_{m \geq 1} {\widetilde k}_m(n) \tA_{n+m}$, 
where the coefficients
${\widetilde k}_m(n)$ are obtained from the corresponding pQCD RGE equations
(with $\A_n \mapsto a^n$, $ \tA_{n+m} \mapsto \ta_{n+m}$). These relations were presented
for any analytic QCD model in Refs.~\cite{GCCV1,GCCV2} in the case of
integer $n$, and in Ref.~\cite{GCAK} for noninteger $n = \nu$. 
The recurrence relations leading to the above relations,
for integer $n$ and within the context of
the minimal analytic (MA) model of Refs.~\cite{ShS,MSS,Sh,BMS}, 
were presented in Refs.~\cite{Shirkov:2006nc,Shirkov}.
Such construction of higher power analogs $\A_n$, not as powers of $\A_1$ 
but rather as linear (in $\A_1$) operations on
$\A_1$, reflects a very desirable functional feature: their compatibility 
with linear integral transformations (such as Fourier or Laplace) 
\cite{Shirkov:1999np}. On the other hand, in linear tranformations, the
image of a power is in general not the power of the image.} 
This means that
the evaluated observables in analytic QCD have the (truncated)
``modified analytic'' (man) series form analogous to the (truncated)
``modified perturbation'' (mpt) series form in pQCD (\ref{Dmpt})-(\ref{Dtmpt})
\bea
{\cal D}(Q^2)_{\rm man} &=& \A_1(\mu^2) + 
\sum_{j=1}^{\infty} {\td}_j(\mu^2/Q^2) \; \tA_{j+1}(\mu^2) \ ,
\label{Dman}
\\
{\cal D}(Q^2;\mu^2)_{\rm man}^{[2 M]} &=& \A_1(\mu^2) + 
\sum_{j=1}^{2 M - 1} {\td}_j(\mu^2/Q^2) \; \tA_{j+1}(\mu^2) \ .
\label{Dtman}
\eea
In view of the presented resummation method (\ref{dBG}), this
is intriguing, because it shows that the series in logarithmic
derivatives of the coupling play a central role both in 
the mentioned resummation method [cf.~Eqs.~(\ref{Dmpt}), (\ref{tDpt})]
and in the evaluation procedure in analytic QCD models 
[Eqs.~(\ref{anrule1})-(\ref{Dtman})].

The reason for the necessity, in the analytic QCD models, 
of the evaluation of the observables via Eq.~(\ref{Dtman}) originates
from the fact that the unphysical renormalization scheme (RS) 
dependence of the truncated series (\ref{Dtman}) is\footnote{
The relation (\ref{RSDtman}) can be obtained in complete analogy
with the perturbative QCD, under the correspondence (\ref{anrule1}).}
\be
\frac{\partial {\cal D}(Q^2;{\rm RS})_{\rm man}^{[N]}}{\partial ({\rm RS})}
= {\widetilde k}_{N}(\mu^2/Q^2) \tA_{N +1}(\mu^2)  + {\cal O}(\tA_{N +2}) \; (\sim \A_{N +1}) \ , 
\qquad ({\rm RS} = \ln \mu^2; c_2; c_3; \ldots) \ ,
\label{RSDtman}
\ee 
and from the fact that in analytic QCD models we have the hierarchy 
$\A_1(\mu^2) > |\tA_2(\mu^2)| > |\tA_3(\mu^2)| \cdots$ at all complex $\mu^2$. We stress that
the expression on the right-hand side of Eq.~(\ref{RSDtman}) contains only
terms $\tA_{j}(\mu^2)$ ($j \geq N+1$) and no other type of terms. For example,
if RS=$\ln \mu^2$, the right-hand side of Eq.~(\ref{RSDtman}) is exactly
$- \beta_0 N {\td}_{N-1}(\mu^2/Q^2) \tA_{N+1}(\mu^2)$.
If we performed
the evaluation by the replacement $a^n \mapsto \A_{1}^n$ ($n \geq 2$),
the resulting truncated analytic power series 
\be
{\cal D}(Q^2;{\rm RS})_{\rm an TPS}^{[2 M]} = \A_1(\mu^2) + 
\sum_{j=1}^{2 M - 1} d_j(\mu^2/Q^2) \A_1(\mu^2)^{j+1} \ .
\label{DanTPS}
\ee
would possess in general an increasingly strong RS dependence when the
order of the truncation $N$ increases
\be
\frac{\partial {\cal D}(Q^2;{\rm RS})_{\rm an TPS}^{[N]}}{\partial ({\rm RS})}
= k_{N}(\mu^2/Q^2) \A_1^{N +1}(\mu^2)  + {\cal O}(\A_1^{N +2}) + {\rm NP}_N \ ,
\label{RSDanTPS}
\ee  
where the terms ${\rm NP}_N$ denote nonperturbative terms ($\sim (\Lambda^2/\mu^2)^k$),
which in general become more complicated and increase in their value 
when $N$ increases. The origin of such terms is the difference 
$\A_1(\mu^2) - a(\mu^2) \sim (\Lambda^2/\mu^2)^m$ at $\mu^2 > \Lambda^2$.

It is evident that our approximant in analytic QCD, Eq.~(\ref{dBGan}),
is RScl invariant (since $\tal_j$ and $\tQ_j^2$ are). Furthermore, in
complete analogy with the pQCD case, we can show that it fulfills
the approximation requirement analogous to Eq.~(\ref{dBGappr})
\be
{\cal D}(Q^2)_{\rm man} - {\cal G}^{[M/M]}_{{\cal D}}(Q^2;{\rm an.}) = {\cal O}(\tA_{2 M+1}) \ ,
\label{dBGanappr}
\ee   
where the right-hand side has only terms of the form $\tA_{j}(Q^2)$ 
($j \geq 2 M + 1$). The relation (\ref{dBGanappr}), together with 
the aforementioned hierarchy of $\tA_j$'s in analytic QCD, gives us
additional hope that our approximants (\ref{dBGan}) will give us
values increasingly close to the full value ${\cal D}(Q^2)_{\rm man}$,
Eq.~(\ref{Dman}), in any chosen analytic QCD model. We will see in the
next Section, on the example of the Adler function at low momenta
($Q^2 = 2 \ {\rm GeV}^2$) that this hope is well grounded.

\section{Numerical checks of the quality of the approximants}
\label{sec:num}
In this Section we will investigate how our approximants (\ref{dBGan})
[and (\ref{dBG})] work when applied to a spacelike QCD observable
whose perturbation series is known to a sufficiently high order.
Specifically, we will consider the massless Adler function
${\cal D}(Q^2)$ at low $Q^2$ ($Q^2 = 2 \ {\rm GeV}^2$) and perform numerical
evaluations of our approximants both in perturbative QCD (pQCD)
and in three different analytic QCD (anQCD) models, namely: 
\begin{itemize}
\item
Minimal Analytic (MA) model of Refs.~\cite{ShS,MSS,Sh}; 
\item
the approximately perturbative anQCD model of Ref.~\cite{CCEM} (CCEM); 
\item
the perturbative anQCD model type ``EE'' (whose beta function involves
exponential functions) in two variants, of Ref.~\cite{GCRKCV}.
\end{itemize}

The characteristics of these different models will be specified in 
more detail later in this Section. Beforehand, we sketch the 
general procedure:
we will consider first the leading-$\beta_0$ (LB) resummation part of 
${\cal D}$ whose expression in pQCD is
\bea
{\cal D}^{\rm (LB)}(Q^2)_{\rm pt} & = &   
\int_0^{\infty} \frac{dt}{t} \; F_{\cal D}(t) a(t Q^2 e^{{\overline {\cal C}}})
\label{DLBint}
\\
& = & a(Q^2) + \td_{1,1} \beta_0 \; {\ta}_{2}(Q^2) +
\cdots +  \td_{n,n} \beta_0^n \; {\ta}_{n+1}(Q^2) + \cdots \ .
\label{DLBexp}
\eea
Here, $F_{\cal D}(t) \equiv w_{\cal D}(t) t$ 
is the characteristic function of the Adler function,
whose explicit expression was obtained in Ref.~\cite{Neubert} on the
basis of the leading-$\beta_0$ expansion coefficients 
$d_{n}^{\rm (LB)} = \td_{n}^{\rm (LB)} = \td_{n,n} \beta_0^n$ obtained from the
leading-$\beta_0$ Borel transform of Refs.~\cite{Ben,Broad} 
at RScl $\mu^2=Q^2$ in the ``V'' scale convention.\footnote{
Ref.~\cite{Neubert} uses the notation ${\widehat D}(t) = 4 F_{\cal D}(t)/t$.
Note that we use throughout the ``${\overline {\rm MS}}$'' 
convention for the scale $\Lambda$, i.e., 
$ {\cal C}={\overline {\cal C}} = -5/3$. Large-$\beta_0$ calculations
are usually performed with ``V'' scale convention, i.e.,  ${\cal C}=0$.
The relations between the two, at a given RScl $\mu^2$ (e.g., $\mu^2=Q^2$), are: 
$\td_{n,n}({\overline \Lambda})= 
\td_{n,n}(\Lambda_{\rm V}) + \sum_{k=1}^{n-1} 
(n!/(k! (n-k)!) (-{\overline {\cal C}})^k \td_{n-k,n-k}(\Lambda_{\rm V}) + 
(-{\overline {\cal C}})^n$.}
The coefficient $\td_{n}^{\rm (LB)}$ represents simultaneously
the leading-$\beta_0$ part
of $\td_n$ and of $d_n$ once these two coefficients are organized in
series in powers of $n_f$ and thus of $\beta_0$; 
$\td_{n}^{\rm (LB)}$ is RSch independent but RScl dependent (see also
Eq.~(\ref{tdnmu}); for details, see Ref.~\cite{GCCV2}).

The evaluations will be performed 
in the simplest renormalization scheme $c_2 = c_3 = \cdots = 0$
in various QCD models (pQCD and anQCD's, except the anQCD model ``EE'').
This is
convenient because the expressions are then simple and explicitly related
with the Lambert function \cite{Gardi:1998qr,Magradze:1998ng}. 
As the point of reference
we take the value $a(M_Z^2,{\overline {\rm MS}}) = 0.119/\pi$. This then corresponds
to the value $a(\mu_{\rm in}^2; n_f=3; c_2=c_3=\cdots=0) \approx 0.2215/\pi$
at the ``initial'' chosen scale $\mu_{\rm in} = 3 m_c= 3.81$ GeV. 

We will assume that $n_f=3$ in our calculations.
At $Q^2 = 2   {\rm GeV}^2$ we obtain $a(2{\rm GeV}^2)=0.3479/\pi$.

The practical evaluations can be performed by choosing any value of
RScl $\mu^2$, e.g. $\mu^2=Q^2$.
In the leading-$\beta_0$ case the choice $\mu^2=Q^2$ 
means using the coefficients 
$\td_{n,n} \equiv \td_{n,n}(\mu^2/Q^2=1)$ in the expansion (\ref{DLBexp}).
Nonetheless, as shown, the use of different RScl $\mu^2 \not= Q^2$ gives us
identical results, as can be checked numerically as well.
We note that by choosing $\mu^2=Q^2$, the coefficients
$d_j \equiv d_j(1)$, $\td_j \equiv \td_j(1)$ are $Q^2$-independent. Therefore,
the weight coefficients $\tal_j$ and parameters $\tu_j$ in 
Eqs.~(\ref{dBG}) and (\ref{dBGan}) are $Q^2$-independent (when $\mu^2=Q^2$), 
and thus the ratio of scales $\tQ_j^2/Q^2 = \exp({\tu}_j/\beta_0)$ 
[see Eq.~(\ref{MMdecom2}), with $\mu^2=Q^2$] will be $Q^2$-independent 
(and, of course, $\mu^2$-independent).
In Table \ref{t1} we give the values of weights $\tal_j$ and scale
ratios $\tQ_j^2/Q^2$ for various indices $M$ of our approximants.
\begin{table}
\caption{The weight coefficients $\tal_j$ and the scale ratios
$\tQ_j^2/Q^2$ for our RScl-invariant approximants, Eqs.~(\ref{dBG})
and (\ref{dBGan}), for various order indices ($M=1,2,3,4$), in the case of
leading-$\beta_0$ massless Adler function.}
\label{t1}  
\begin{ruledtabular}
\begin{tabular}{lllll}
$M$ & ($\tal_1$; $\tQ_1^2/Q^2$) & ($\tal_2$,$\tQ_2^2/Q^2$) & ($\tal_3$,$\tQ_3^2/Q^2$) &
($\tal_4$,$\tQ_4^2/Q^2$)
\\ 
\hline
$M=1$ & (1; 0.5001) & - & - & - 
\\
$M=2$ & (0.6948; 0.1711) & (0.3052; 5.771) & - & -
\\
$M=3$ & (0.3579; 0.07969) & (0.6011; 1.0534) & (0.0410; 85.77) & -
\\
$M=4$ & (0.1376; 0.03803) & (0.6821; 0.3862) & (0.1767; 17.16) & (0.0037; 1518.)
\end{tabular}
\end{ruledtabular}
\end{table}
We can see from the Table that the scale ratios $\tQ_j^2/Q^2$ get increasingly
spread out when the order index $M$ increases. However, for those ratios
which are much smaller or much larger than unity, the corresponding weight 
factors are small.

The authors of Ref.~\cite{BEGKS} applied the diagonal Pad\'e approximants 
to the (auxiliary) power series quantity 
${\widetilde {\cal D}}^{\rm (LB)}(Q^2)_{\rm pt}$
(at $Q^2 = 2 \ {\rm GeV}^2$) obtained from the series (\ref{DLBexp})
by the replacement ${\ta}_{n+1} \mapsto a^{n+1}$ (the approximation of 
one-loop RGE running), and compared with the result of the integration
(\ref{DLBint}) obtained by assuming one-loop RGE running of
$a(t Q^2 e^{{\overline {\cal C}}})$; the integral is ambiguous
in the integration at low $t$ (IR regime) due to the
Landau singularity, so they chose the Principal Value for the integration.

The results of this type of (one-loop) evaluation are given in Table \ref{t2},
for the case $Q^2 = 2 \ {\rm GeV}^2$. We fix the one-loop running
coupling $a_{1\ell}(Q^{' 2})$ so that it agrees with the
aforementioned full $a$ at $Q^2=2 \ {\rm GeV}^2$:
$a_{1\ell}(Q^2) = a(Q^2) =  0.3479/\pi$.
In addition, we include in the Table the corresponding results 
with the full pQCD evaluation in the $c_2=c_3= \cdots 0$ renormalization
scheme (``two-loop'') which uses in the integral (\ref{DLBint}) the full
pQCD $a(t Q^2 e^{{\overline {\cal C}}})$, and our approximants (\ref{dBG}).
\begin{table}
\caption{The results of the one-loop approach: diagonal Pad\'e 
approximants (dPA) ([M/M]) with increasing index $M$, at
$Q^2=2 \ {\rm GeV}^2$, for the leading-$\beta_0$ massless Adler function
${\cal D}(Q^2)$.
For comparison, the result of the
Principal Value of integration (with the estimated IR renormalon ambiguity)
is included. In addition, the approximants (\ref{dBG}) in the case of
full pQCD running $a(t Q^2 e^{{\overline {\cal C}}})$ are included, and
the corresponding Principal Value. 
In the parentheses, the corresponding results of the truncated series
(\ref{DLBexp}) are given (with RScl $\mu^2=Q^2$). See the text for details.}
\label{t2}  
\begin{ruledtabular}
\begin{tabular}{cllllll|l}
Case  & $M=1$ & $M=2$ & $M=3$ & $M=4$ & $M=5$ & $M=6$ & PV
\\ 
\hline
1-loop  &  0.134(0.130) & 0.161(0.155) & 0.175(0.164) & 0.194(0.160) & -0.497(0.080) & 0.156(-0.714) &
$0.178 \pm 0.020$
\\
full     &  0.140(0.134) & 0.200(0.174) & 0.532(0.198) & $0.095\!-\!0.051i$(0.107) & $0.162\!-\!0.009i$ (-1.79)&
$0.250\!-\!0.001i$ (-39.8)& $0.174 \pm 0.020$
\end{tabular}
\end{ruledtabular}
\end{table}
We can see that the dPA's (in the one-loop case) 
and our approximants (\ref{dBG})
oscillate rather erratically around the corresponding Principal Value. This
has to do with the fact that, at higher order index $M$ ($M \geq 3$)
the scales $\tQ_j^2$ come rather close to the Landau singularity of
the running perturbative coupling. In fact, the approximants
become even complex in the full case once at least one of the scales
$\tQ_j^2$ hits the unphysical cut $(0, \Lambda^2_{\rm L.})$ (where:
$\Lambda^2_{\rm L.} \approx 0.150 \ {\rm GeV}^2$, i.e., $\Lambda_{\rm L.} \approx
.388$ GeV), since $a(\tQ_j^2)$ becomes
complex. In the one-loop case, we have a simple Landau pole instead of the cut
(with $\Lambda^2_{\rm L.} \approx 0.036 \ {\rm GeV}^2$, i.e., 
$\Lambda_{\rm L.} \approx0.190$ GeV), so the approximants would remain real even
when one of the scales were below the Landau pole. In the parentheses, the
results of the corresponding truncated series are given -- for the one-loop
case the truncated version ${\widetilde {\cal D}}(Q^2;\mu^2)_{\rm pt}^{[2 M]}$ of
the expansion (\ref{tDpt}), and in the full (loop) case the truncated
version ${\cal D}(Q^2;\mu^2)_{\rm mpt}^{[2 M]}$ Eq.~(\ref{Dtmpt}), both with RScl
$\mu^2=Q^2$. We see that these truncated series behave in general
worse than the resummed versions, and show for larger $M$ asymptotic
divergence (in the one-loop case for $M \geq 5$, and in he full loop case for
$M \geq 4$). 

There are several analytic QCD models (for $\A_1(Q^2)$) in the literature.
The most used one is the model of Shirkov, 
Solovtsov and Milton \cite{ShS,MSS,Sh},
which keeps for the cut of $\A_1(Q^2)$ on the negative $Q^2$ axis 
the discontinuity function of the pQCD coupling $a(Q^2)$, 
and the unphysical pQCD cut on the positive axis 
is eliminated
\be
\A_1^{\rm (MA)}(Q^2) = \frac{1}{\pi} \int_0^{\infty} d \sigma 
\frac{\rho_1^{\rm (pt)}(\sigma)}{\sigma + Q^2} \ ,
\label{MA}
\ee
where ${\rho}_1^{\rm (pt)}(\sigma) = {\rm Im} a(Q^{' 2} = - \sigma - i \epsilon)$.
This represents, in a sense, the minimal changes (in the cut) 
with respect to pQCD. Therefore, we call this model the
Minimal Analytic (MA).\footnote{In the literature, it is usually
called Analytic Perturbation Theory (APT), and it then involves
a specific construction of the analytic analogs of higher powers
$a^n$. The construction can be applied only in MA, and it is in such a case
equivalent to the construction presented in Refs.~\cite{GCCV1,GCCV2}
(the latter construction being applicable to any anQCD).} 
The only adjustable parameter there is the scale ${\overline \Lambda}$  
(in the ``${\overline {\rm MS}}$'' scale convention). 
In order to reproduce QCD phenomenology at high energies, 
the value of this scale at $n_f=5$ 
in MA is about $260$ MeV, which corresponds at $n_f=3$
to the value of ${\overline \Lambda} \approx 415$ MeV \cite{Bakulev}.
We will use this value in MA, and will use there also the RSch
$c_2=c_3=\cdots = 0$.

Another analytic QCD model is described in Ref.~\cite{CCEM} (CCEM). It differs
from MA in the sense that the discontinuity function
$\rho_1(\sigma) = {\rm Im} \A_1(-\sigma - i \epsilon)$ differs from the
pQCD discontinuity function at low $\sigma \alt 1 \ {\rm GeV}^2$
where it is replaced by a delta function. 
The spacelike coupling $\A_1$ is then
\be
\A_1(Q^2) = \frac{f_1^2}{u + s_1} + \frac{1}{\pi}
\int_{s_0}^{\infty} ds \frac{r_1^{\rm (pt)}(s)}{s + u} \ ,
\label{CCEM}
\ee
where $u \equiv Q^2/\Lambda_{\rm W}^2$, $s \equiv \sigma/\Lambda_{\rm W}^2$,
$r_1^{\rm (pt)}(s) \equiv \rho_1^{\rm (pt)}(\sigma)$ (in the RSch $c_2=c_3= \cdots=0$), 
and $\Lambda_{\rm W} \approx 0.487$ GeV is the scale appearing in the 
Lambert function $W_{\mp 1}(z_{\pm})$. The scale $\Lambda_{\rm W}$ was 
fixed basically by the requirement
that the high energy QCD phenomenology be reproduced.
The (dimensionless) free parameters ($f_1^2$, $s_1 \equiv M_1^2/\Lambda_{\rm L}^2$, 
$s_0 \equiv M_0^2/\Lambda_{\rm L}^2$) are
fixed in the model in such a way that at high $Q^2$ it merges with the pQCD 
coupling to a high degree of accuracy [$\A_1(Q^2) - a(Q^2) \sim (\Lambda^2/Q^2)^3$]
and that, simultaneously, it reproduces the measured value of the
semihadronic (massless and strangeless)
tau decay ratio\footnote{
We use the variant of the model with the value of $s_0=3.858$, 
which reproduces the measured
value of $r_{\tau}$ when the leading-$\beta_0$ resummation and the inclusion
of the known beyond-the-leading-$\beta_0$ terms is performed in the
evaluation of $r_{\tau}$.}
$r_{\tau}(\triangle S=0, m_q=0)_{\rm exp.} = 0.203 \pm 0.004$.
We note that in MA we have 
$\A_1^{\rm (MA)}(Q^2) - a(Q^2) \sim (\Lambda^2/Q^2)$, i.e., at high energies
this difference is not quite negligible, and the predicted value
of  $r_{\tau}(\triangle S=0, m_q=0)$ is about 0.14.

Yet another analytic QCD model which we will use is the so called
EE model of Ref.~\cite{GCRKCV}, which is in fact a fully perturbative
analytic QCD model [the $\beta(a)$ function is analytic function of 
$\A_1(Q^2) \equiv a(Q^2)$ at $a=0$].\footnote{
Our general construction of $\A_n(Q^2)$ gives in such models:
$\A_n = \A_1^n$, as it should be.} 
The beta function has the Ansatz
\be
\beta(a) = - \beta_0 a^2 (1 - Y) f(Y)|_{Y \equiv a/a_0} \ ,
\label{betans}
\ee
where $a_0 = a(Q^2=0)$ is a finite value (infrared fixed point),
$f(Y)$ is analytic at $Y=0$, and we require analyticity of $a(Q^2)$ at
$Q^2=0$, which turns out to give the condition $a_0 \beta_0 f(1) = 1$. The
expansion of $\beta(a)$ in powers of $a$ also has to reproduce the
first two universal coefficients $\beta_0$ and $\beta_1$, cf.~Eq.~(\ref{RGE}).
There are at least two variants of the mentioned "EE'' model. In the first
variant (``EEv1'') the function $f(Y)$ in the beta function is a combination
of (rescaled and translated) functions $(e^Y - 1)/Y$ and $Y/(e^Y-1)$
$(e^Y-1)/Y$ and $Y/(e^Y-1)$:
\be
{\rm EEv1:} \qquad
f(Y) =  \frac{ \left( \exp[- k_1 (Y - Y_1)] -1 \right) }
{ [ k_1 (Y - Y_1) ] }
\frac{ [ k_2 (Y - Y_2) ] }{ \left( \exp[- k_2 (Y - Y_2)] -1 \right) }
\times {\cal K}(k_1,Y_1,k_2,Y_2) \ ,
\label{EE}
\ee 
where the constant ${\cal K}$ ensures the required normalization
$f(Y=0)=1$. In this variant we have, at first, five real parameters:
$a_0 \equiv a(Q^2=0)$ and $Y_j, k_j$ ($j=1,2$). Two parameters ($Y_2$ and $a_0$)
are eliminated by the aforementioned conditions: $a_0 \beta_0 f(1) = 1$ and
the reproduction of the universal $\beta_1$ coefficient. The other three
parameters are approximately fixed by the condition of analyticity of
$a(Q^2)$ and the requirement of obtaining as high a value of 
$r_{\tau}(\triangle S=0, m_q=0)$ as possible (it is always too low in
comparison to the experimental value $0.203 \pm 0.004$). The obtained values
are: $Y_1=0.1$, $k_1=10.0$, $k_2=11.0$. This results in $a_0=0.236$ and the highest
possible value $r_{\tau}(\triangle S=0, m_q=0) \approx 0.15$. This latter value 
is still clearly too low.

The second version ("EEv2'') has the function $f(Y)$ in the beta function
modified, in comparison to EEv1, by a factor $f_{\rm fact}$
\bea
{\rm EEv2:} \qquad f_{\rm EEv2}(Y) &=& f_{\rm EEv1}(Y) f_{\rm fact}(Y) \ ,
\label{EEv2f}
\\
{\rm with} \   f_{\rm fact}(Y) &=& \frac{(1 + B Y^2)}{(1 + (B+K) Y^2)}
\quad (1 \ll K \ll B) \ .
\label{EEv2ffact}
\eea
This factor has the values of $K$ and $B$ adjusted so that the
expansion of the evaluation of $r_{\tau}(\triangle S=0, m_q=0)$, by the inclusion
of the leading-$\beta_0$ (LB) contribution and of the first three 
beyond-the-leading-$\beta_0$ (bLB) contributions, gives the correct
$r_{\tau}$ value: $r_{\tau}(\triangle S=0, m_q=0) = 0.203$
($ \Rightarrow $ $B=1000$ and $K=5.4$). The factor
$f_{\rm fact}(Y)$ does not destroy the analyticity of $a(Q^2)$, and
it does not change substantially the values of $a(Q^2)$ since it is
close to the value one for most $Y$'s. However,
the price that we pay is high nonetheless: the coefficients
$c_j \equiv \beta_j/\beta_0$ of the expansion of the modified beta function 
are extremely high for $j \geq 4$ ($c_j \agt 10^6$ for $j \geq 4$), implying
strong divergence of any evaluation series of observables (including $r_{\tau}$)
when bLB terms of $\sim a^n$ with $n \geq 5$ are included.
The factor $f_{\rm fact}(Y)$ introduces singularities of
$\beta(a)$ at rather small values of $|a|$.

For more details on the models CCEM (with $s_0=3.858$) and EEv1 and EEv2,
we refer to Refs.~\cite{CCEM} and \cite{GCRKCV}, respectively.

The results of our approximants (\ref{dBGan}) in
these analytic QCD models, for the leading-$\beta_0$ part
of the Adler function ${\cal D}(Q^2)$ at $Q^2=2 \ {\rm GeV}^2$, 
are presented in Table \ref{t3}.
For comparison, the exact integrated values
 \be
{\cal D}^{\rm (LB)}_{\rm an}(Q^2) =
\int_0^{\infty} \frac{dt}{t} F_{\cal D}(t) \A_1(t Q^2 e^{{\overline {\cal C}}})
\label{DLBintan}
\ee
are also given in the Table. Note that the leading-$\beta_0$ integration,
Eq.~(\ref{DLBintan}), has now no ambiguities since no Landau singularities
exist, in contrast to the pQCD case (\ref{DLBint}).
Incidentally, the expansion of Eq.~(\ref{DLBintan}) is completely analogous
to the pQCD expansion (\ref{DLBexp})
\be
{\cal D}^{\rm (LB)}(Q^2)_{\rm man}
= \A_1(Q^2) + \td_{1,1} \beta_0 {\tA}_{2}(Q^2) +
\cdots +  \td_{n,n} \beta_0^n {\tA}_{n+1}(Q^2) + \cdots \ .
\label{DLBman}
\ee
\begin{table}
\caption{Evaluations of the leading-$\beta_0$ massless Adler function
${\cal D}^{\rm (LB)}(Q^2)$ in various analytic QCD models, using our RScl 
invariant approximants
(\ref{dBGan}), with increasing index $M$, at $Q^2=2 \ {\rm GeV}^2$.
For comparison, the exact result of the integration (\ref{DLBintan})
is included. In parentheses in the Table, the values of the corresponding
truncated series ${\cal D}^{\rm (LB)}(Q^2)^{[2 M]}_{\rm man}$ are given (with RScl
$\mu^2=Q^2$). See the text for details.}
\label{t3}  
\begin{ruledtabular}
\begin{tabular}{clllllll|l}
model  & $M=1$ & $M=2$ & $M=3$ & $M=4$ & $M=5$ & $M=6$ & $M=7$ & exact
\\ 
\hline
MA  &  0.1167(0.1147) & 0.1222(0.1214) & 0.1217(0.1208) & 0.1217(0.1205) & 0.1217 (0.1211)& 0.1217(0.1209) &
0.1217(0.1174) & 0.1217
\\
CCEM     &  0.1371(0.1321) & 0.1649(0.1640) & 0.1650(0.1733) & 0.1617(0.1788) & 0.1624(-0.0048) & 0.1632(-0.0407)  & 0.1626(11.70) & 0.1627
\\
EEv1     & 0.1062(0.1047) & 0.1141(0.1144) & 0.1136(0.1146) & 0.1131(0.1138) & 0.1132(0.1063) & 0.1133(0.0842) & 0.1133(12.42) & 0.1133
\\
EEv2     & 0.0965(0.0952) & 0.1036(0.1035) & 0.1035(0.1041) & 0.1032(0.1041) & 0.1032(0.1018) & 0.1032(0.0840) & 0.1032(-0.4615) & 0.1032
\end{tabular}
\end{ruledtabular}
\end{table}
In parentheses, we give the results of the corresponding truncated
version of the series (\ref{DLBman}), i.e., 
${\cal D}^{\rm (LB)}(Q^2)^{[2 M]}_{\rm man}$, with $\mu^2=Q^2$, for each $M$.
We see in the Table that our approximants converge systematically and fast
to the exact values when the order index $M$ increases. The truncated series,
on the other hand, have divergent behavior which, though, starts
manifesting itself at large $M$'s ($M \geq 7$ in the MA case; $M \geq 5$ in the
CCEM and EE cases) since these are analytic QCD models. Despite this divergence,
the aforementioned hierarchy of the couplings
$|\tA_{k}(Q^2)| > |\tA_{k+1}(Q^2)|$ in general
turns out to be true for all relevant indices $k$ in the Table
($k=1,\ldots,13$), at $Q^2 = 2 \ {\rm GeV}^2$. 

The first three coefficients $d_j$ ($j=1,2,3$) are now exactly known for
the Adler function \cite{d1,d2,d3}. 
Therefore, we can construct our approximants
(\ref{dBGan}) for the order indices $M=1$ and $M=2$ on the basis of these
exact four coefficients. The results of this calculation, for the
three analytic QCD models, are presented in Table \ref{t4}. For comparison,
we also include the results (tman) of the truncated modified analytic
series (\ref{Dtman}), with $\mu^2=Q^2$, 
and the more refined ``LB+bLB'' evaluation which
takes into account the leading-$\beta_0$ resummation contribution
(\ref{DLBintan}) and the three additional known terms (bLB:
beyond-the-leading-$\beta_0$)
\be
{\cal D}^{\rm (bLB)}(Q^2)_{\rm man}
= \sum_{n=1}^3 (\td_n - \td_{n,n} \beta_0^n) \; {\tA}_{n+1}(Q^2) \ .
\label{DbLBman}
\ee
\begin{table}
\caption{Evaluations of the full massless Adler function
in various analytic QCD models, using our RScl invariant approximants
(\ref{dBGan}) for $M=1,2$, at $Q^2=2 \ {\rm GeV}^2$.
For comparison, two other evaluations 
(LB+bLB; and tman: truncated modified analytic series) 
are included. See the text for details.}
\label{t4}  
\begin{ruledtabular}
\begin{tabular}{cllll}
model  & $M=1$ & $M=2$ & LB+bLB &tman 
\\ 
\hline
MA  & 0.1175 & 0.1196 & 0.1191  &  0.1199
\\
CCEM     &  0.1389 & 0.1535 & 0.1528 & 0.1541
\\
EEv1     & 0.1070 & 0.1164 & 0.1183 & 0.1195
\\
EEv2     & 0.0972 & 0.1390 & 0.1584 & 0.1587
\end{tabular}
\end{ruledtabular}
\end{table}
We can see that our approximants (\ref{dBGan}), with index $M=2$,
represent a competitive evaluation of the observable, especially
when comparing with the (partially) resummed results LB+bLB
and the truncated (modified) analytic series (tman). 

The results of our method with $M=2$, 
in the MA and CCEM cases, deviate from
the LB+bLB results less than the tman results deviate. Since the analytic
models MA and CCEM are in ``tame'' RSch's [i.e., the ones where the
RSch parameters $c_j$ ($j \geq 2$) are very small, in fact, zero],
we can expect that both the LB+bLB and tman approaches
give good estimates of the true value in the model, and
that LB+bLB is probably a better approach since it uses significantly
more input information than tman. However, we recall that our
$M=2$ approximants use as little input information as the truncated (tman)
approach, i.e., the first three $d_j$'s, and yet Table \ref{t4}
indicates that our approximants with $M=2$ are
competitive with the LB+bLB approach in the MA and CCEM models.
 
On the other hand, the RSch coefficients $c_j$ are increasing fast in 
the models EEv1 and dramatically fast in EEv2. In that case, the coefficients 
$\td_j$ and $(\td_j - \td_{j,j} \beta_0^j)$,
which depend on $c_j$ via an additive term $-c_j/(j-1)$ (if $j \geq 2$), increase 
very fast when $j$ increases, so that
tman and LB+bLB approaches become uncertain.\footnote{
The $c_j \equiv \beta_j/\beta_0$ coefficients in EEv2 are: $-106.8 (j=2)$; 
$326.7 (j=3)$; $1.72 \cdot 10^6 (j=4)$; $3.08 \cdot 10^6 (j=5)$, etc.}
We notice that in the case of EEv2,
our approximant (for $M=2$) is essentially different from the LB+bLB and from 
the tman result. The tman series 
(\ref{Dtman}) and the truncated bLB series (\ref{DbLBman}) become
in that case very divergent once we include the terms $\tA_{n+1}$ with
$n\geq 4$ (cf.~Ref.~\cite{GCRKCV} for further details on the
divergence of the coefficients $\td_n$ in this case). In that case,
our approximants, for $M=2$, are probably comparatively the most reliable
estimate of the true result in the EEv2 model. 

\section{Conclusions}
\label{sec:concl}

We tested in various analytic QCD models an earlier developed
\cite{GC,GCRK} RScl invariant resummation method, by applying it
to the evaluation of the massless Adler function
${\cal D}(Q^2)$ at low energy ($Q^2=2 \ {\rm GeV}^2$). The method is global,
i.e., nonpolynomial in the (analytic) coupling parameter. It is related 
with the method of diagonal Pad\'e approximants (dPA's), representing an
extension of the dPA method by achieving exact RScl independence. The
method, applied to spacelike observables, results in a linear combination
of coupling parameters at several spacelike momentum scales 
(each of them RScl invariant),
and thus represents an extension of the well-known scale-setting techniques
of Stevenson \cite{PMS}, Grunberg \cite{ECH}, and 
Brodsky-Lepage-Mackenzie \cite{BLM}. For observables with
low scale $Q^2$ of the process, the method when applied within the
perturbative QCD is not very efficient in practice. The reason for this
is that the perturbative QCD coupling $a(Q^2)$ has unphysical (Landau)
singularities at low positive $Q^2$, and some of the scales of our approximant
turn out to be close or even within this singularity sector. On the other
hand, the method turns out to be very efficient in analytic QCD models,
because the analytic coupling $\A_1(Q^2)$ has no unphysical singularities.
In the case of the leading-$\beta_0$ part of the Adler function, the
results of the method converge very fast to the exact result within
each analytic QCD model. Furthermore, when the method is applied
to the truncated (analytic) series of the entire Adler function, whose
first three coefficients beyond the leading order are known exactly,
the result of the method becomes competitive with the result of the
sum of the (exact) leading-$\beta_0$ (LB) contribution and the truncated
beyond-the-leading-$\beta_0$ (bLB) analytic series, although the latter
method (LB+bLB) uses significanly more input information than our method.
We conclude that our method is at the moment probably the best
method, in the analytic QCD frameworks, for the evaluation of spacelike 
observables when the evaluation is based on the known part of the 
truncated integer power perturbation series of the observable.
The method can be used also for the evaluation of
timelike observables (such as the cross section of
$e^+e^-$ scattering into hadrons, and semihadronic $\tau$ decay ratio $r_{\tau}$)
when the latter are expressed as contour integrals
involving spacelike observables.

\begin{acknowledgments}
\noindent
The authors wish to thank D.~V.~Shirkov for valuable comments.  
This work was supported in part by FONDECYT Grant No.~1095196 
and Anillos Project ACT119 (G.C.),
and DFG-CONICYT Bilateral Project 060-2008 (G.C. and R.K.).
\end{acknowledgments}

\appendix

\section{The approximation requirement}
\label{app}

Here we show that the approximation requirement, Eq.~(\ref{dBGappr}),
is fulfilled by our approximant (\ref{dBG}). Taylor-expanding $a(\tQ_j^2)$'s
in the approximant around $\ln(\mu^2)$, by using the definitions (\ref{tan}),
we obtain
\bea
{\cal G}_{\cal D}^{[M/M]}(Q^2) &= &\sum_{j=1}^M \tal_j a(\tQ_j^2)
= \sum_{j=1}^M \tal_j \sum_{k=0}^{\infty} {\ta}_{k+1}(\mu^2) 
\left(-\beta_0  \ln({\tQ}_j^2/\mu^2) \right)^k
\label{exp1}
\\ 
& = & \sum_{k=0}^{\infty} {\ta}_{k+1}(\mu^2) \sum_{j=1}^M 
\tal_j  \left(-\beta_0  \ln({\tQ}_j^2/\mu^2) \right)^k
=  \sum_{k=0}^{\infty} {\ta}_{k+1}(\mu^2) \sum_{j=1}^M \tal_j (- \tu_j)^k \ .
\label{exp2}
\eea
In the last equation we used the fact that $\tu_j = \beta_0 \ln({\tQ}_j^2/\mu^2)$,
see.~Eqs.~(\ref{MMdecom1})-(\ref{MMdecom2}). However, 
Eqs.~(\ref{MM})-(\ref{MMdecom2}) and (\ref{tDpt}) tell us that
\be
{\widetilde {\cal D}}(Q^2)_{\rm pt} - 
[M/M]_{{\widetilde {\cal D}}}\left( a_{1\ell}(\mu^2) \right) = 
{\cal O}\left( a_{1\ell}(\mu^2)^{2 M+1} \right) \ .
\label{tDdPAappr}
\ee
This implies that the expansion of $[M/M]_{{\widetilde {\cal D}}}(x)$
in powers of $x=a_{1\ell}(\mu^2)$ reproduces\footnote{
This is the expansion of the expression (\ref{MMdecom1}) in powers of $x$.}
the coefficients at powers
of $x^n$ for $n=1,\ldots,2M$ in the expansion of ${\widetilde {\cal D}}(Q^2)$,
Eq.~(\ref{tDpt})
\be
\sum_{j=1}^M \tal_j (- \tu_j)^k = \td_k(\mu^2/Q^2) \qquad
{\rm for} \ k=0,1,\ldots,2M-1 \ .
\label{expintu}
\ee
Note that $\td_0(\mu^2/Q^2) \equiv 1$. Inserting the indentities (\ref{expintu})
into Eq.~(\ref{exp2}), we obtain
\be
{\cal G}_{\cal D}^{[M/M]}(Q^2) = \sum_{k=0}^{2 M - 1} 
{\ta}_{k+1}(\mu^2) \td_k(\mu^2/Q^2) + {\cal O}(\ta_{2M+1}) \ .
\label{dBGexp}
\ee
This, in combination with the expansion (\ref{Dmpt}) of the observable
${\cal D}(Q^2)$ in ${\ta}_{k+1}(\mu^2)$, gives us immediately
\be
{\cal D}(Q^2)_{\rm mpt} - {\cal G}_{\cal D}^{[M/M]}(Q^2) =
{\cal O}( \ta_{2 M + 1} ) = {\cal O}(a^{2 M + 1}) \ ,
\label{dBGapprapp}
\ee
i.e., the approximation identity (\ref{dBGappr}). The same proof can be
repeated in analytic QCD models (except for the notational change
$\ta_{k+1} \mapsto \tA_{k+1}$), i.e., the approximation identity
(\ref{dBGanappr}) is also valid.

\end{document}